\documentclass[9pt,twocolumn,twoside]{opticajnl}
\journal{opticajournal} 

\usepackage{lineno}
\usepackage{xr}
\usepackage{amsmath}

\title{Anomalous vortex beam driven harmonic generation}

\author[1,2,3]{B. Kumar Das}
\author[1,2,3]{M. Ciappina}
\author[4]{W. Gao}
\author[4*]{C. Granados}

\affil[1]{Department of Physics, Guangdong Technion - Israel Institute of Technology, 241 Daxue Road, Shantou, Guangdong, China, 515063}
\affil[2]{Technion - Israel Institute of Technology, Haifa, 32000, Israel}
\affil[3]{Guangdong Provincial Key Laboratory of Materials and Technologies for Energy Conversion, Guangdong Technion - Israel Institute of Technology, 241 Daxue Road, Shantou, Guangdong, China, 515063}
\affil[4]{Eastern Institute of Technology, Ningbo, 315200, China}
\affil[*]{cagrabu@eitech.edu.cn}

\begin{abstract}
The generation and control of the properties of light beams carrying orbital angular momentum is fundamental to extend our understanding on the light-matter interaction process. In this letter, we investigate the use of anomalous and modified anomalous vortex beams for the generation of high-order harmonics (HHG) of the fundamental field. We demonstrate that by controlling the order and topological charge (TC) of the driving field, one can control the vortex beam size of the generated harmonics. A key outcome of this control is the ability to drive the HHG process with fundamental beams of higher TCs and consequently generating harmonics with higher TCs ($\approx 100$) while maintaining a compact beam size and nearly uniform divergence in the far-field across a wide range of harmonic orders.  
\end{abstract}

\setboolean{displaycopyright}{false} 

\begin{document}

\maketitle

High-order harmonic generation (HHG) \cite{AttoPhys,AttoSci} is a highly nonlinear, non-perturbative process that enables the generation of extreme ultraviolet (XUV) light when an ultrashort laser pulse interacts with a target medium \cite{Lewenstein2}. A key characteristic of the generated harmonics is their ability to be synthesized into either a train of pulses or a single pulse with attosecond duration \cite{Lewenstein1}. This advancement has not only opened new avenues for investigating fundamental physical processes but also led to various industrial applications \cite{Wang17, Li2018}. 

In recent years, HHG driven by structured light—specifically, light carrying orbital angular momentum (OAM)—has enabled the generation of twisted XUV \cite{Geneaux,CarlosPRL} even with pulse durations in the attosecond domain \cite{AttoVortex}. Both fundamental and harmonic fields are characterized by the topological charge (TC) which describes the number of twists the light wavefront undergoes around the propagation axis per wavelength \cite{VorBeam}. These beams are famously called spatial vortex (SV) beams. In addition, light beams can also carry OAM in the direction transverse to the direction of propagation and are called spatiotemporal optical vortex (STOV) beams  \cite{STOV2012}. Generally, both SV and STOV beams can be described by an amplitude and a phase term. The amplitude term is defined in terms of different polynomials and special functions e.g., Laguerre-Gaussian (LG) Beams and Bessel-Gaussian (BG) Beams \cite{VorBeam}. The phase term $\exp(-il\phi)$ is called the helical phase where $l$ is the TC carried by the beams, and $\phi$ is the azimuthal angle. The later describes the coupling of the spatial variables $\arctan(y/x)$ in the case of the SV beams or the spatiotemporal variables $\arctan(t/x)$ in the case of STOV beams. 

A fundamental characteristic of the HHG process driven by either SV or STOV beams, is that the OAM is conserved \cite{CarlosPRL}. More importantly, this conservation leads to harmonic radiation with a total TC per photon that scales with the harmonic order, $l_q=ql$, where $q$ is the harmonic order and $l$ is the TC of the fundamental driving field. However, the value of the TC directly determines the divergent behavior of the far-field intensity profile of the vortex beam, making it difficult to control and characterize \cite{HTC}. Additionally, it has been demonstrated that most of the characteristics of the fundamental field are imprinted into the harmonic field \cite{BPOV, ELGB}. Vortex beams have found multiple applications ranging from trapping particles to telecommunications \cite{Vappl1,Vappl2,Vappl3}, making it fundamental to understand its generation and propagation characteristics. Additionally, harmonic fields carrying OAM can potentially be used to investigate processes sensitive to the beam topology as for example the spin-orbit process in solids and the detection of chiral molecules \cite{Chiral1,Chiral2}. 

Anomalous vortex beams possess many important characteristics: exhibiting fractional OAM, distortion, self-healing properties, or non-trivial topologies \cite{PropAnomalous,MAVB,HollowG}. All of these properties are still to be applied and exploited in the harmonic generation process. One important beam, holding great potential for driving non-linear process, is the anomalous vortex (AV) beam with self-focusing properties \cite{PropAnomalous}. In Ref.~\cite{PropAnomalous}, the authors experimentally demonstrated the generation of AVB using a spatial light modulator. Additionally, they discussed the propagation of AVB in a series of optical elements using the ABCD matrix formalism. Self-focusing is a fundamental mechanism in maintaining the vortex beam integrity and its propagation over long distances \cite{SelfF}. It means that self-focusing helps to counteract diffraction. Particularly in HHG driven by structured light beams \cite{BPOV,CarlosPRL,HHGAiry}, self-focusing could enhance the local intensity, improving the efficiency of attosecond pulse generation. Added to this, AV beams are typically used as virtual sources for the generation of elegant vortices which stand alone for its application to scenarios beyond the paraxial approximation \cite{ELGB}. What sets them apart is a complex argument that symmetrically links the polynomial and Gaussian components, resulting in propagation dynamics that are both mathematically elegant and physically coherent. An important effect of the TC on the beam size of conventional vortex beams is the change of the beam size: an increase in the TC leads to the beam size expansion, decreasing the overall intensity. It could significantly affect the harmonic generation process because of its nonlinear character (sensitive to the fundamental field's intensity). 

We recently demonstrated that one way to avoid such problem is to drive the harmonic generation process by a perfect optical vortex (POV) beam with beam size independent of the TC, resulting in different harmonic vortices sharing similar divergence in the far-field. Another way to mitigate this is by using modified AV (MAV) beams with controllable intensity distribution and beam size. Additionally, the AV and MAV beams offer the flexibility of reducing the beam size and producing Bessel and elegant Laguerre-Gauss beams. Importantly, these characteristics are translated to the harmonic generation process, as will be shown.

\begin{center}
\begin{figure}[h!]
\includegraphics[width=0.44\textwidth]{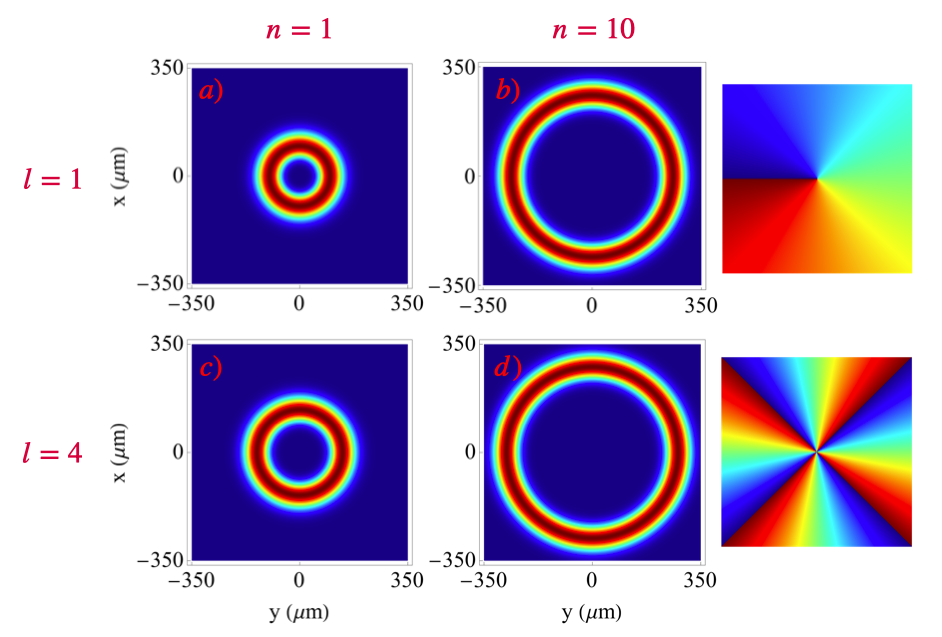}
\caption{Normalized intensity distribution for the fundamental anomalous vortex beam. In different panels, we present the changes in the intensity distributions for different values of $n$ and $l$.  In the phase structures (and for all the plots), the minimum (blue color) and  maximum (red color) correspond to -$\pi$, and $\pi$, respectively. For all plots, $\delta=0$.}\label{Fig1} 
\end{figure} 
\end{center}
\begin{center}
\begin{figure}[h!]
\includegraphics[width=0.44\textwidth]{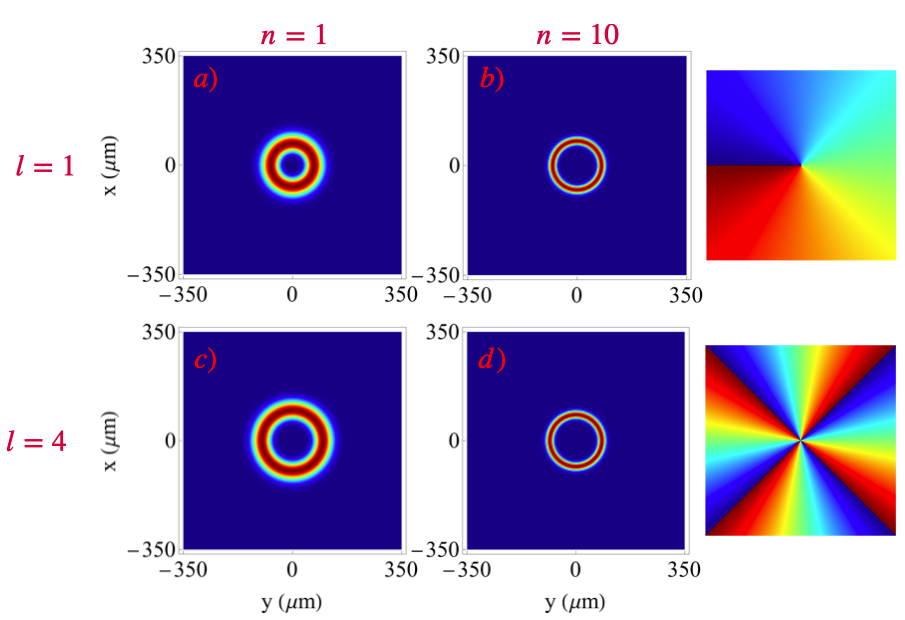}
\caption{Normalized intensity distribution for the fundamental anomalous vortex beam. In different panels, we present the changes in the intensity distributions for different values of $n$ and $l$. For all plots, $\delta=1$.}\label{Fig2} 
\end{figure} 
\end{center}

We start by describing the electromagnetic field corresponding to a MAV beam which encapsulates the field distribution of an AV beam within it. The spatial complex amplitude describing both MAV and AV beams is given by: 

\begin{eqnarray}
    E_{M} (\hat{\rho},\phi,z=0) =  \hat{E}_0\left( \frac{\hat{\rho}}{w_0} \right)^{(2n+|l|)}\exp{\left( -\frac{\hat{\rho}^2}{w_0^2} \right)} \exp{\left(-i l \phi\right)} \label{eq1}
\end{eqnarray}

with $\hat{\rho} = \sqrt{(\delta n+1)}\rho$, and phase $ \phi = \arctan\left(y/x\right)$. The beam waist width is represented by $w_0=80$~$\mu$m. The constants $\delta$ and $n$ represent the modification parameter with a range $[0,1]$ that affect the self-focusing properties of the beam and the order of the beam, respectively~\cite{MAVB}. The case of $\delta=0$ corresponds to the traditional AV beam. With $n=\delta=0$ Eq.~\ref{eq1} reduces to the LG beam with zero radial index. Here, $\hat{E}_0=\sqrt{2^{| l| +2 n+1}(\delta  n+1)P_0/(\pi w_0^2 (2n+|l|)!)}$ with $P_0$ being the input power. In Fig.~\ref{Fig1}, we show the intensity and phase profiles of the fundamental AV beam for different values of $n$ and $l$ (Eq.~\ref{eq1}) with $\delta = 0$. For these beam parameters, the fundamental beam exhibits a large beam size for large TC values and for large values of the order $n$ through the relation $w=w_0 \sqrt{2n+|l|+1}/\sqrt{\delta n+1}$, where $w$ is the second moment width. This is a typical behavior found in conventional vortex beams and have a fundamental problem: One can not imprint a large TC in the fundamental beam since large beam size reduces the peak intensity of the beam which in turn affects the HHG process. Consequently, controlling the beam size while maintaining the maximum intensity in the beam is fundamental. We can exploit the parameter $\delta$ to maintain the fundamental beam's small spot size. It can be seen from the equation, $\hat{\rho} = (\delta n+1)\rho$, that larger values of $\delta$ corresponds to larger values of $\hat{\rho}$ and consequently smaller beam size. We present the intensity and phase profiles of the MAV beam for different values of $l$ and $n$ for $\delta=1$ in Fig.~\ref{Fig2}. It can be seen from the figure that increasing the beam order leads to a reduction in beam size, with higher-order MAV beams showing narrower ring widths compared to their lower-order counterparts. It can also be seen from the Fig.~\ref{Fig2} that the spot size of the MAV beam exhibits small sensitivity to the changes in TCs for higher values of the beam order. This result is extremely important since it allow us to conclude that we can drive the HHG process with MAV beams with large TCs. This result is similar to the scenario presented by the perfect optical vortex \cite{BPOV}.  
\vspace{-1cm}
\vspace{-0.3cm}

\noindent  It is important to calculate the radius of the maximum intensity as a function of the TC since it quantifies how the divergence of the beam depends on the TC. For this, we calculate $\partial I(\rho,z)/\partial \rho=0$. For the conditions presented here, we locate the gas-target at the focal point of the fundamental field, i. e., at $z=0$. The calculation leads to the radius of maximum intensity: $\rho_{max}^{MAV} = w_0\sqrt{(2n+|l|)/2}\times1/(\delta n+1)$. Interestingly, $\rho_{max}$ reduces to $\rho_{max}^{LG}=\omega_0\sqrt{|l|/2}$) for $n=0$ and $\delta=0$ which corresponds to the radius of the maximum intensity for LG beams \cite{BPOV}. For other values, it is clear that one can compensate the action of the TC (increasing the beam size) with the modification parameter, $\delta$. Notice also that the numerator of $\rho_{max}^{MAV}$ increases $\sqrt{2n+1}$
but the denominator increases with ($\delta n+1$). This translates into a slower increase in $\rho_{max}^{MAV}$ for large values of $n$. In conclusion, one can control the action of the TC on the radius of maximum intensity by use of the beam order and the modification parameter.

\noindent The HHG process in atomic gases driven by AV and MAV beams can be simulated using the thin-slab model (TSM)\cite{Hernandez2015}, which consists in calculating the far-field harmonic amplitude and phase using the near-field harmonic amplitude and phase combined with the Fraunhofer diffraction integral. In the TSM, the near-field harmonic amplitude is proportional to the fundamental beam's amplitude raised to the power $p$, where, $p$ is the scaling factor that is less than the harmonic order $q$ in case of plateau harmonics. It is important to highlight that the scaling factor $p$ is similar for harmonics in the plateau region ($p\approx 3$), while $p=q$ for the perturbative region of the harmonic spectrum.
In addition, the near-field phase scales with $q$ times the phase of the fundamental beam plus the dipole phase. It is important to highlight that the TSM is valid for the scenarios where the dipole approximation is valid. This means, the electron excursion in the continuum of energy is negligible compared to the driving field wavelength. Additionally, the TSM in its current form is applicable only when the paraxial approximation is followed. Using the TSM, the far-field complex amplitude of the $q^{th}$ order harmonic can be written as:  

\begin{eqnarray}
    E^{far}_{MAV} (\beta,\theta) &=& \left(\frac{C}{\tau^{j}}\right)^{3/2} E_0^p i^{l q} \pi ^{l q+1} \exp{\left(- i \Phi_{d}\right)} \exp{\left(-iql\theta\right)}   \nonumber \\
    &\times& \left(\frac{p(1+\delta n)}{w_0^2}\right)^{-m} \left(\frac{1+\delta n}{w_0}\right)^{p (l+2 n)} \Gamma \left[m\right]   \nonumber \\
    &\times& _1F_1 \left[ m, l q+1 , -\frac{\pi^2 w_0^2 \tan^2(\beta)}{p \lambda_q^2(1+\delta n )^2} \right] \left( \frac{\tan^2(\beta)}{\lambda_q^2} \right)^{\frac{l q}{2}}, \label{FF}
\end{eqnarray}

\noindent here, $m=n p+\frac{1}{2} l (p+q)+1$, $\Phi_{d} = - \alpha_{q}^{j}|E_0|^2$ corresponds to the dipole phase, and $C$ is a constant. $\tau^{j}$, and $\alpha_{q}^{j}$ denote the excursion time and a strong-field parameter associated with $j^{th}$(short or long) quantum path, respectively. $\lambda_q=\lambda/q$ represents the wavelength of the $q^{th}$ order harmonic. Furthermore, $\beta=\sqrt{\beta_{x}^2+\beta_{y}^2}$ and $\theta$ represent the divergence and azimuthal angle in the far field, respectively. In our simulation, we use the fundamental beam wavelength $\lambda = 800$~nm and $p=3$ ($I =0.87 \text{ to }1.7\times10^{14}$ W/cm$^2$). Notice that for calculating Eq.~\ref{FF}, we use only short trajectories and we have neglected the dipole phase contribution. The dipole phase gives rise to secondary OAM contributions of around one order of magnitude smaller than the helical phase \cite{NPtwist}. 

\begin{center}
\begin{figure}[h!]
\includegraphics[width=0.44\textwidth]{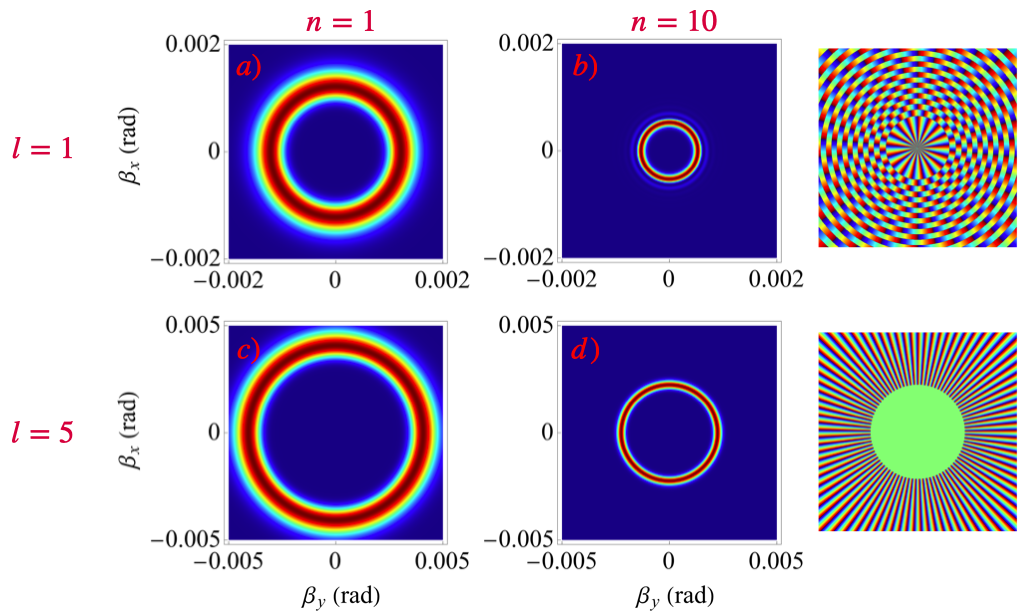}
\caption{Normalized intensity distributions for the 17$^{\text{th}}$ harmonic. In different panels, we present the changes in the intensity distributions for different values of $n$ and $l$. For all plots, we use $\delta=0$.}\label{Fig3} 
\end{figure} 
\end{center}
\vspace{-.5cm}

\noindent In Fig.~\ref{Fig3} (a) and (c), we present the far-field intensity distributions of the harmonic order $17^{\text{th}}$ for $\delta=0$, $n=1$, and TC values $l=1$ and $l=5$, respectively. In panels (b) and (d), we show the far-field intensity distributions of the same harmonic order but for $n=10$. The far-field phase plots corresponding to $n=10$, $l=1$, and $n=10$, $l=5$, are shown in the right hand side of Fig.~\ref{Fig3}. From the far-field phase profiles, the law of TC upscaling i.e., $l_{q}=ql$ can be clearly verified by counting the number of $2\pi$ phase shifts along the azimuthal coordinate of the harmonic vortices. A remarkable difference is observed for different values of $l$: while for $l=1$ and $n=10$, we see multiple rings around the vortex core (see panel (b)), there is no ring around the vortex core for $l=5$ and $n=10$ (see panel (d)). This is also reflected in the phase plots presented in the right hand side of Fig.~\ref{Fig3}. This is an important characteristic since it demonstrates that for lower TC values, the harmonic field preserves the main characteristic of the elegant vortex beam \cite{MAVB}. It is important to note that the phase plot reflects the presence of multiple rings in the intensity distribution for panel (b). However, the intensity of the outer rings are extremely smaller as compared to the visible ring, therefore, cannot be observed clearly in the panel (b). For large values of TC, a single, thin-ring intensity distribution is observed instead of multiple rings as shown in panel (b). The control over the beam size and the external rings of the vortex beams by exploiting the TC and the beam order manifest the anomalous behavior of the beams presented here. Also notice that for a fixed value of $n$, an enhancement in the TC value reduces the external rings in the vortex beam (Fig.~\ref{Fig3} (b) and (d)). Additionally, the resulting intensity distribution characterized by external rings makes the AVB a candidate to generate Bessel and elegant vortex beams \cite{MAVB} with large TC in the XUV regime. This represents a clear extension of its capability to generate Bessel and elegant beams beyond the visible region of the electromagnetic spectrum \cite{MAVB}. 

\vspace{-.3cm}

\begin{center}
\begin{figure}[h!]
\includegraphics[width=0.44\textwidth]{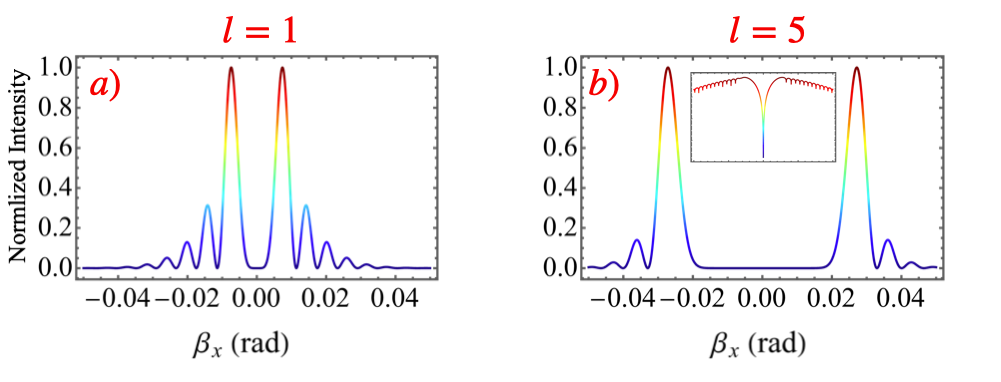}
\caption{Normalized line intensity profiles for $\beta_y=0$. The inset, in logarithmic scale, shows oscillations beyond $\beta_x=0.05$~rad.}\label{Fig4} 
\end{figure} 
\end{center}

\vspace{-.4cm}

\noindent The control over the number of rings around the singularity of the vortex beam with the TC was corroborated by calculating the intensity distribution of $3^{\text{rd}}$ harmonic at the far field. Notice that we chose this harmonic since higher harmonics will not exhibit a large number of rings because of their large TC which additionally demonstrate the control over the number of rings. The results are presented in Fig.~\ref{Fig4}. For all the panels in Fig.~\ref{Fig4}, we use $n=10$ and $\delta = 1 $. In panels (a) and (b), we show the normalized line intensity profiles for $\beta_y=0$ for TC values $l=1$ and 5. In particular, the inset in panel (b) shown in logarithmic scale, demonstrate the oscillatory behavior of the intensity distribution beyond $\beta_x=0.05$~rad. This clearly indicates the generation of Bessel and elegant vortex beams. For the calculation of the $3^{\text{rd}}$ harmonic, we use the perturbative scaling law, see e.g. Ref.~\cite{Hernandez2015}. 

\begin{center}
\begin{figure}[h!]
\includegraphics[width=0.44\textwidth]{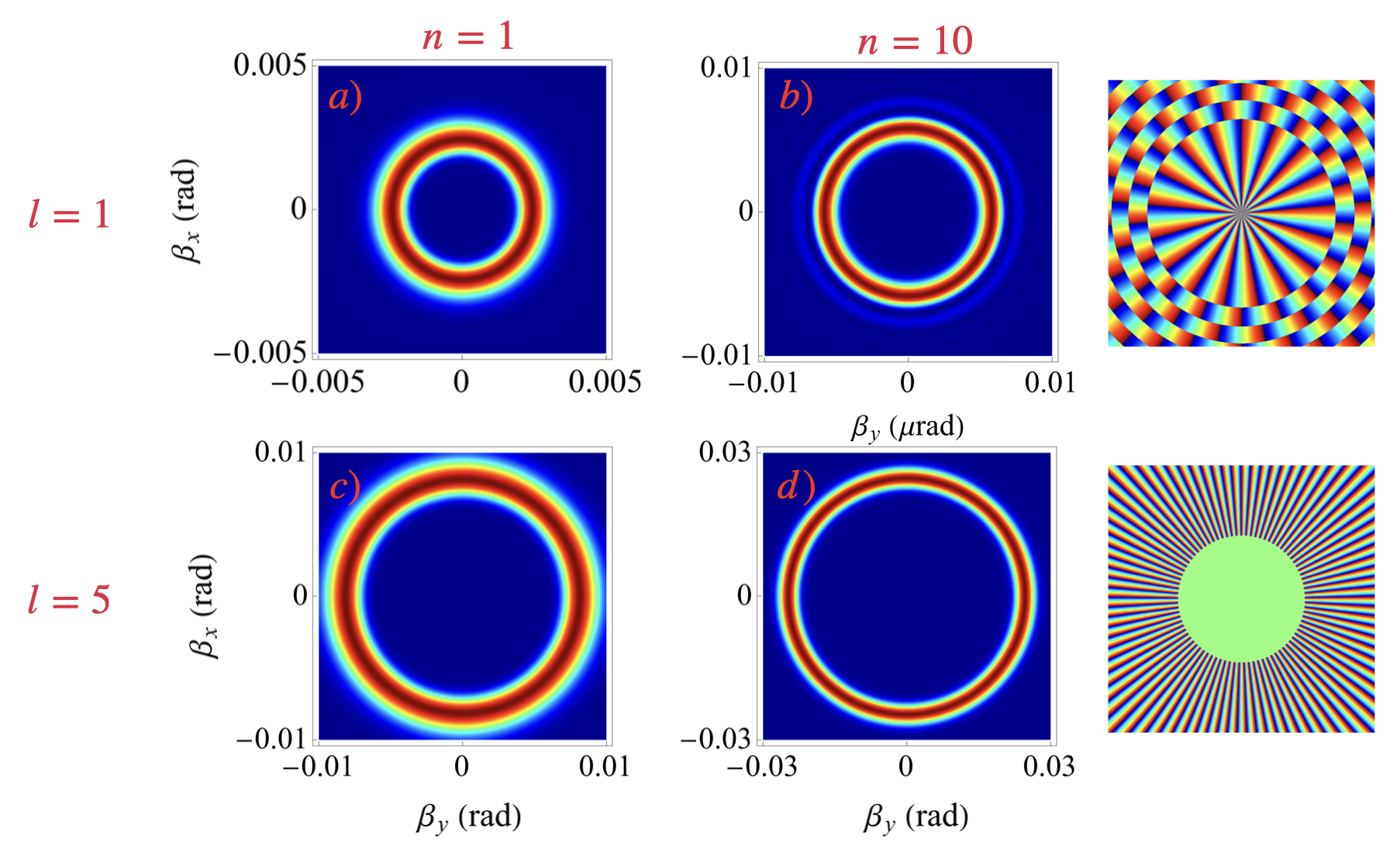}
\caption{Normalized intensity distributions for the 17$^{\text{th}}$ harmonic. In different panels, we present the changes in the intensity distributions for different values of $n$ and $l$. For all plots, we use $\delta=1$.}\label{Fig5} 
\end{figure} 
\end{center}

\vspace{-1.5cm}

\begin{center}
\begin{figure}[h!]
\includegraphics[width=0.44\textwidth]{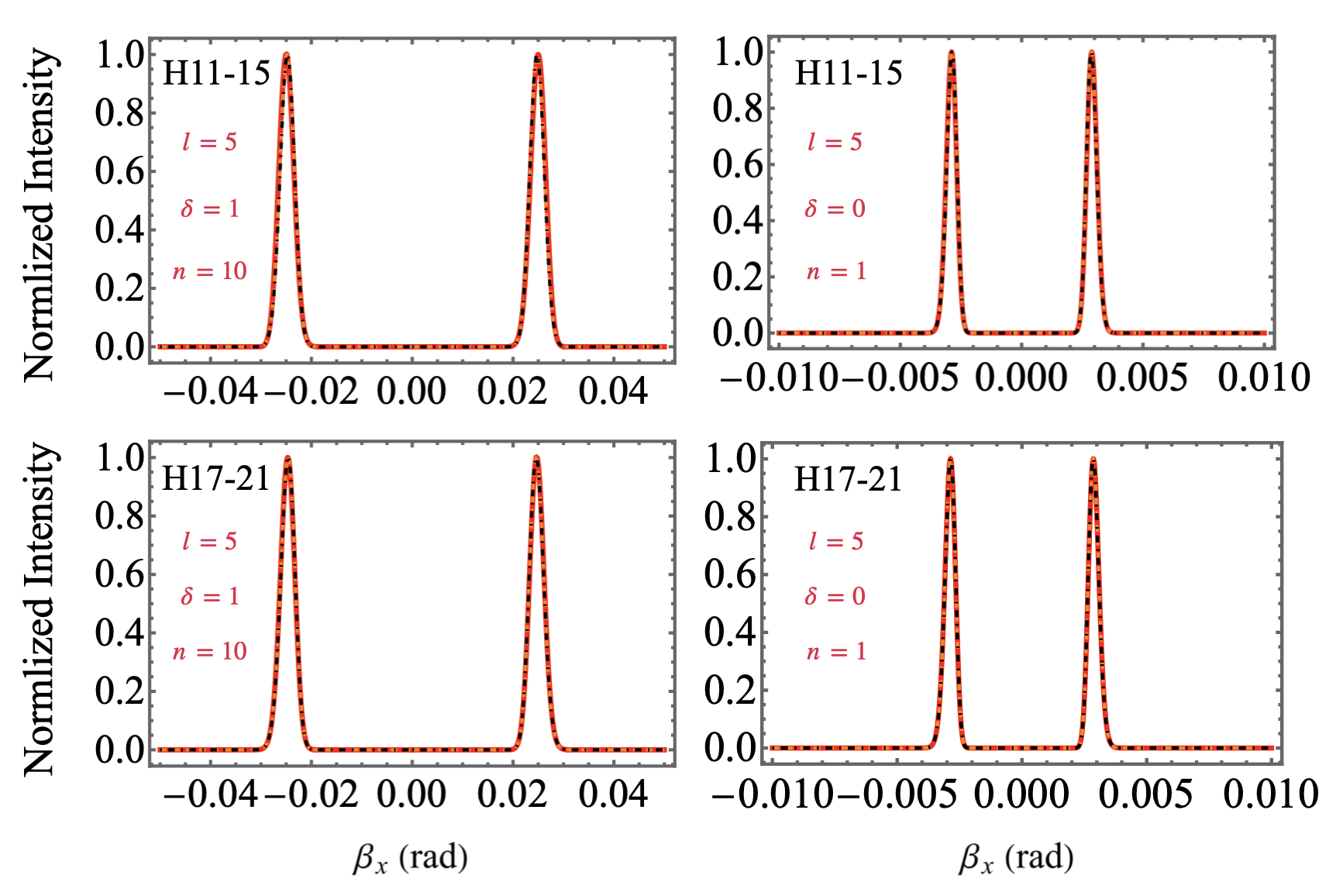}
\caption{Normalized intensity distribution for different harmonic orders. }\label{Fig6} 
\end{figure} 
\end{center}
\vspace{-.5cm}

\noindent We further investigate the effect of changing the modification parameter ($\delta$) in the HHG process. The results are shown in Fig.~\ref{Fig5}. It is important to note that external rings also appear in the intensity distributions for the $\delta=1$ case like $\delta=0$ case. However, an important difference can be observed between the two cases ($\delta=0$ and 1): the harmonic vortex beam size increases as $\delta$ increases its value for large TCs. This is opposite to the fundamental beam behavior shown in Figs.~\ref{Fig1} and~\ref{Fig2}. In addition, the phase plots shown at the right side of panels (b) and (d), are identical to those presented in Fig.~\ref{Fig3}, which is consistent with the conservation of OAM. The phase plots show that for lower values of the TC, the harmonic field exhibits elegance in the far-field, while for large values of TC the elegance is lost.

\noindent It is extremely important to understand how the size of the harmonic vortices changes in the far-field. If similar beam size is maintained for different harmonics, there is a possibility towards generating twisted attosecond pulse train or attosecond light spring. For this end, we calculate the normalized line intensity profiles for harmonic orders 11$^{\text{th}}$ to 21$^{\text{st}}$ for different values of $\delta$. The results are presented in Fig.~\ref{Fig6}. For this calculation, we use the same $p$-value for different harmonic orders and observe that there is no such significant changes in the intensity distribution when different $p$-values are used for different harmonic orders. From Fig.~\ref{Fig6}, it is clear that the divergence is same for different harmonics considered in this case and changes in the value of $\delta$ or $n$ doesn't affect this behavior.

In conclusion, we investigated the generation of high-order harmonics in atomic gases driven by AV and MAV beams. We have demonstrated control over the beam size of the fundamental field and, consequently, over the generated harmonics. Additionally, we have shown that in the far-field, for different values of the fundamental beam order and the self-focusing parameter, harmonic vortices can exhibit elegance and, therefore, can be used to generate elegant vortex beams in the XUV regime. Notably, we examined the 17$^{\text{th}}$ harmonic order, which corresponds to a wavelength of approximately 47~nm. For a fundamental beam of TC $l=5$, the TC of the 17$^{\text{th}}$ harmonic is $l_{17}=85$ due to the conservation of OAM in the HHG process. Likewise, for the harmonic order  21$^{\text{st}}$, the resulting TC value is $l_{21}=105$, demonstrating the potential to produce controllable XUV beams with higher TCs. Notably, we found that all considered harmonics exhibit the same divergence, indicating a robust and scalable approach for generating high-TC XUV beams. 
\begin{backmatter}
\bmsection{Disclosures} The authors declare no conflicts of interest.
\bmsection{Acknowledgments} B. K. Das and M. F. Ciappina acknowledge the National Key Research and Development Program of China (Grant No.~2023YFA1407100), Guangdong Province Science and Technology Major Project (Future functional materials under extreme conditions - 2021B0301030005) and the Guangdong Natural Science Foundation (General Program project No. 2023A1515010871).
\end{backmatter}
\vspace{-0.2cm}
\bibliography{literatur}

\bibliographyfullrefs{literatur}

\end{document}